\documentclass[aps,llncs,twocolumn]{revtex4-1} 
\usepackage{graphicx}
\usepackage{amsmath}
\usepackage{float}
\begin{document}
\title{Spin-orbit driven spin depolarization in the ferromagnetic Weyl semimetal Co$_3$Sn$_2$S$_2$ } 

\author{Sandeep Howlader$^1$}

\author{Surabhi Saha$^2$}
\author{Ritesh Kumar$^1$}
\author{Vipin Nagpal$^3$}

\author{Satyabrata Patnaik$^3$}
\author{Tanmoy Das$^2$}
\author{Goutam Sheet$^1$}
\email{goutam@iisermohali.ac.in}
\affiliation{$^1$Department of Physical Sciences, 
Indian Institute of Science Education and Research Mohali, 
Mohali, Punjab, India}

\affiliation{$^2$ Department of Physics, Indian Institute of Science, Bengaluru, Karnataka, India}
\affiliation{$^3$ School of Physical Sciences, Jawaharlal Nehru University, Delhi, India}

\begin{abstract}

Co$_3$Sn$_2$S$_2$ has recently emerged as a ferromagnetic Weyl semi-metal. Theoretical investigation of the spin-split bands predicted half metalicity in the compound. Here, we report the detection of  a spin polarized supercurrent through a Nb/Co$_3$Sn$_2$S$_2$ point contact where Andreev reflection is seen to be large indicating a large deviation from half metallicity. In fact, analysis of the Andreev reflection spectra reveals only 50\% spin polarization at the Fermi level of Co$_3$Sn$_2$S$_2$. Our theoretical calculations of electronic Density of States (DOS) reveal a spin depolarizing effect near the Fermi energy when the role of spin-orbit coupling is included. Inclusion of spin-orbit coupling also reveals particle-hole asymmetry that explains a large asymmetry observed in our experimental Andreev reflection spectra. 

\end{abstract}
\maketitle


 Co$_3$Sn$_2$S$_2$  has recently attracted significant attention due to presence of a unique ferromagnetic Weyl semimetallic phase\cite{Wan, Xu, Fang, Shi, Weng, Huang, XuS, Chu} with possible half metallicity in the compound. The ferromagnetic phase was experimentally measured to appear below 177 K. The band structure calculations on Co$_3$Sn$_2$S$_2$ reveals existence of ferromagnetic order along with a band gap for spin minority band at the Fermi level \cite{Schnelle,Kubodera, Holder}.   Support for half metallicity was also obtained from angle-resolved and core-level photo-emission spectroscopy\cite{Wang} on Co$_3$Sn$_2$S$_2$. Recently the compound has come under scanner again because several experiments indicated a novel topological Weyl phase in Co$_3$Sn$_2$S$_2$. In this context, a large intrinsic anomolous Hall effect\cite{Xu2} was measured in crystals of Co$_3$Sn$_2$S$_2$, where the Weyl fermions provided dominant contribution to the Hall conductivity. Details of the topological surface Fermi arc states were investigated theoretically\cite{Groot,Venkatesan,Zutic11,yan}  in Co$_3$Sn$_2$S$_2$. A giant anomalous Nernst signal\cite{Yang}, much larger than the known ferromagnets, was obtained on the material where it was argued that the enhanced contribution was due to a Berry curvature\cite{Lu,Haldane} close to the Fermi level. The unique combination of topological Weyl behavior and high spin-polarization may open up new areas of Weyl-spintronics. Therefore, investigation of the spin-resolved transport properties of the Fermi surface of Co$_3$Sn$_2$S$_2$ is of utmost importance. While the spin polarization of the spin-split bands of the system was earlier investigated by photoemission spectroscopy, the transport spin polarization of the Fermi surface remains an outstanding problem.

In this Letter, we report measurement of transport spin polarization in Co$_3$Sn$_2$S$_2$ by spin-resolved point contact Andreev reflection(PCAR) spectroscopy\cite{Upadhyay,Auth, Kamboj, aggarwal, Anshu, Mukhopadhyay}. Our measurements show that contrary to the expectation within band structure calculations, the transport spin polarization in Co$_3$Sn$_2$S$_2$ is $\sim50$\%, not 100\%. Detailed analysis of the experimental data and our calculations including the role of spin-orbit coupling in the compound reveal a prominent spin depolarization effect near the Fermi energy driven by spin-orbit coupling and thus the significantly reduced transport spin polarization is explained.

High quality single crystals of Co$_3$Sn$_2$S$_2$ were used for the low temperature experiments presented here. Co$_3$Sn$_2$S$_2$ single crystals were synthesized using modified Bridgman method\cite{Holder,Vaqueiro}. For polycrystalline Co$_3$Sn$_2$S$_2$, stoichiometric amounts of Co lumps (Sigma Aldrich, 99.999\%), Sn powder (Sigma Aldrich, 99.999\%) and S powder were uniformly mixed in an agate mortar and pelletized. The mixture was then transferred to a vacuum evacuated silica tube. The tube was kept in the furnace for two periods of 48 hrs at 500$^o$C and 700$^o$C with heating and cooling rate of 0.5$^o$C/min in both the cycles. An intermediate grinding of the samples were to be used. To grow single crystals, the single phase Co$_3$Sn$_2$S$_2$ polycrystalline powder was taken in an alumina crucible and again vacuum sealed in a quartz tube. The tube was then slowly heated to 1000$^o$C over 30 hrs, maintained at 1000$^o$C for 24 hrs and slowly cooled to 800$^o$C within 72 hrs. Afterwards, the furnace was turned off and the sample was air quenched to room temperature. Shiny crystals were obtained after cleaving the as-grown ingot with a razor blade. The crystal structure and phase of the single crystals were determined from the Reitveld refinement of powder X-ray diffraction. The refinement confirms the trigonal crystal structure with spacegroup ${R\bar{3}m(166)}$. The derived lattice parameters are  $a$ = 5.3686 \AA and $c$ = 13.1747 \AA. The resistivity measurement of Co$_3$Sn$_2$S$_2$ revealed metallic behaviour in the whole temperature range with a small kink observed at 177 K signalling a magnetic transition. The resistivity $\rho$(2K)is found to be 14 $\mu\Omega$-cm with RRR = 35 which  implies good quality of our crystals. Next, we have performed temperature dependent measurement of magnetization $M$ with $H$ = 500 G. Transition to the ferromagnetic phase is again confirmed from the measurement. The derivative of magnetization shows the sharp magnetic transition around 177K. 

\begin{figure}
	\centering
		\includegraphics[width=.5\textwidth]{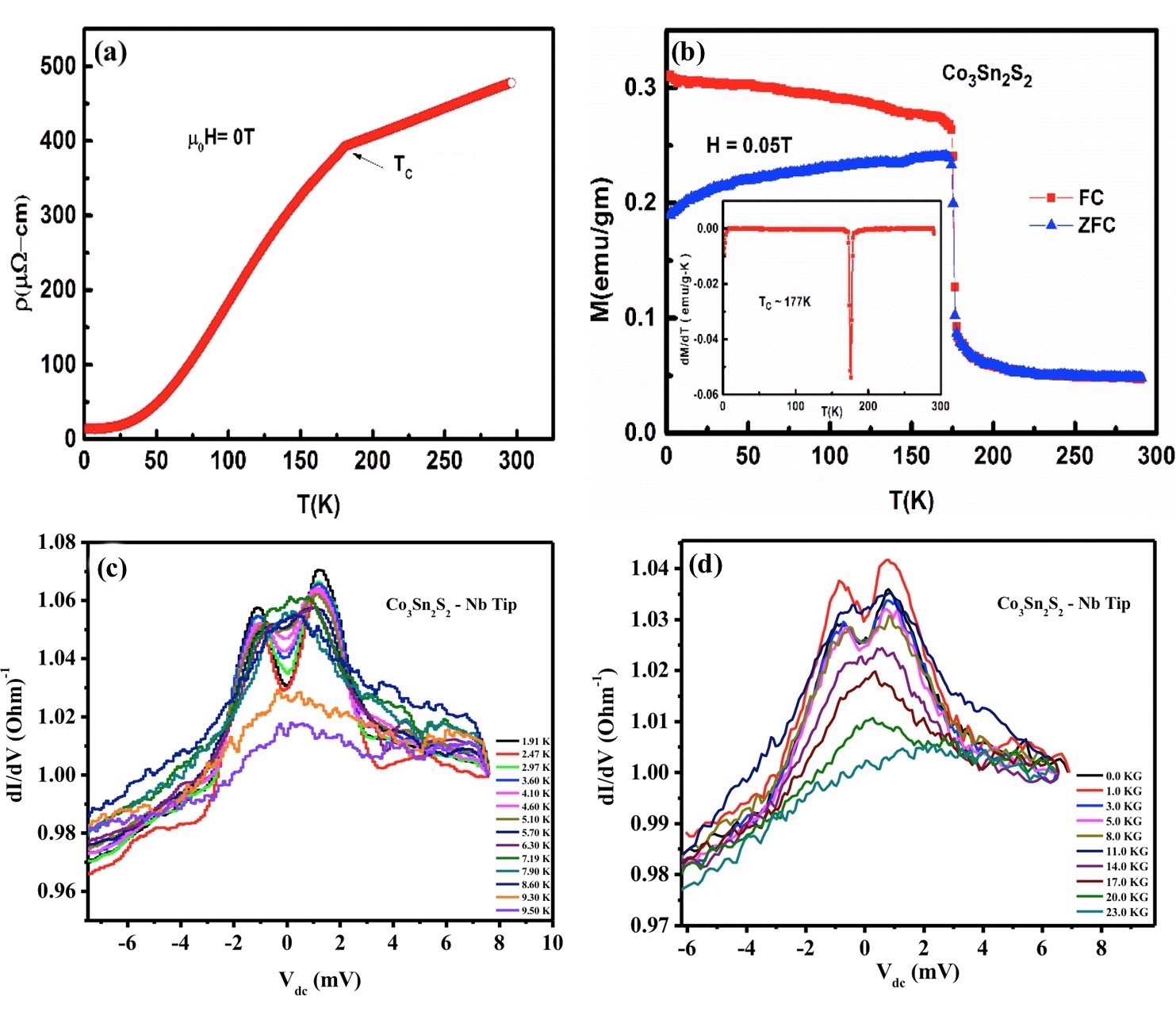}\\
		\caption{(a)Resistivity Vs temperature plot showing ferromagnetic to paramagnetic transition at Curie temperature $T_c$=177 K (b) Variation of magnetization with temperature showing  field cool and zero field cool curves (c) Temperature dependent assymetric PCAR spectra (d) Magnetic field dependent assymetric PCAR spectra.}

\end{figure}

All the measurements were done in a liquid helium cryostat equipped with a superconducting vector magnet (6T-1T-1T). For temperature dependent experiments, a variable temperature insert (VTI) was used which operates between 1.4 K and 300 K. The point contact spectroscopy\cite{Duif,Sheet,Jansen} experiments were performed using a home-built probe based on the standard needle-anvil method. The point contacts were formed $in-situ$ at low temperatures. We have carefully chosen the ballistic point contacts for analysis and data with critical current dominated artefacts have been identified and rejected.
\begin{figure} [h!]
	\centering
	\includegraphics[width=0.5\textwidth]{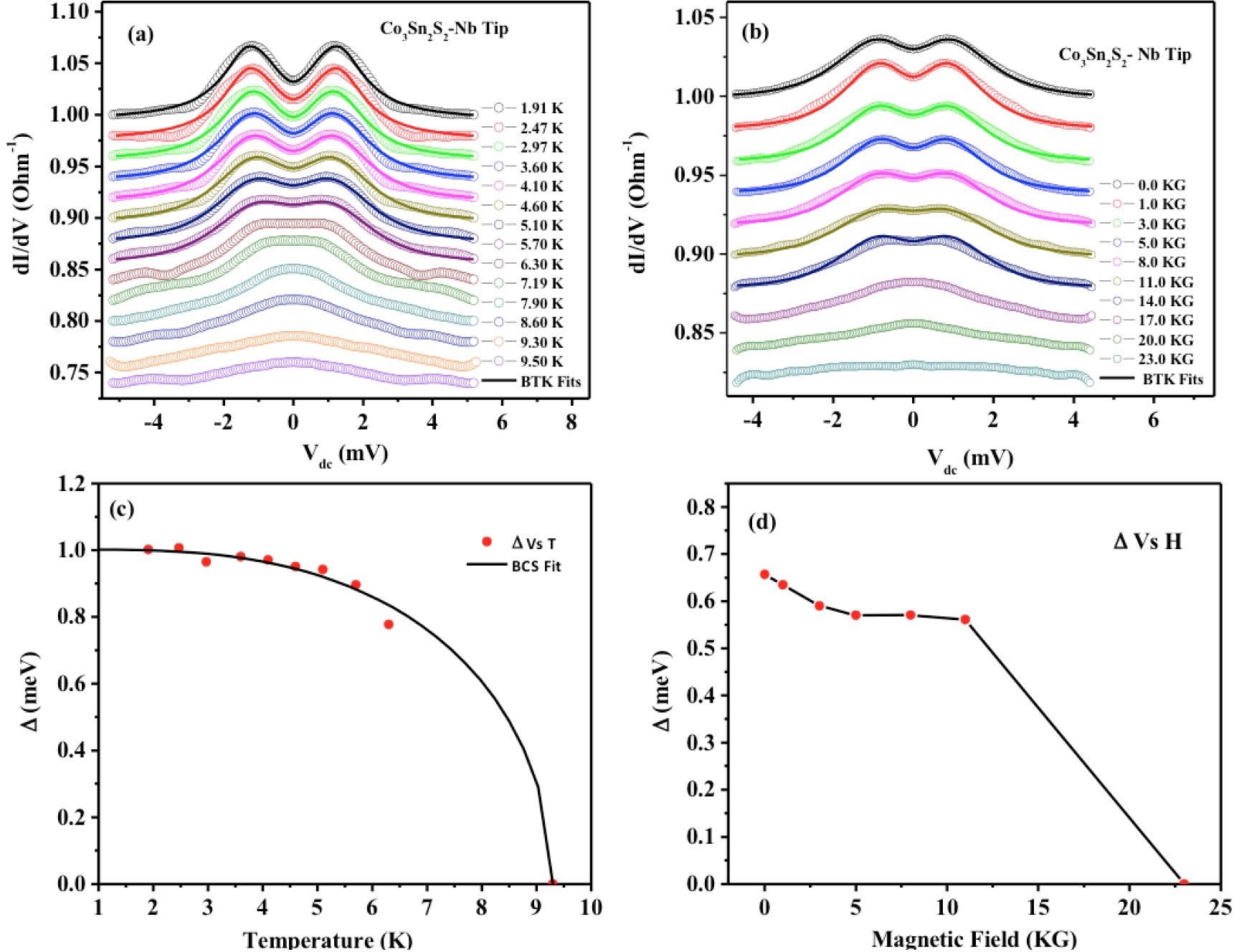}     
		\caption{ (a) Temperature dependence of conductance spectra with corresponding BTK fits (b) Magnetic field dependence of conductance spectra with corresponding BTK fits (c) $\Delta$ Vs T with BCS fit (d) $\Delta$ Vs H}	

\end{figure}
The electronic transport through a ballistic point-contact between a normal metal and a conventional superconductor is dominated by a quantum process called Andreev reflection that involves the reflection of a spin up (down)
electron as a spin down (up) hole from the interface. Andreev reflection leads to a special kind of non-linearity in the $I-V$ spectrum which is directly probed in a $dI/dV$ vs. $V$ spectrum recorded across such a point contact. The
Andreev reflection spectra are traditionally analyzed by a model developed by Blonder, Tinkham, and
Klapwijk (BTK)\cite{Blonder, Strijkers}. The model assumes a $\delta$-function
potential barrier whose strength is characterized by a dimensionless parameter $Z$ which is proportional to the strength of the barrier at the interface. Ideally, for elemental superconductors where the
quasi-particle life-time (usually represented by $\Gamma$) is very large, the Andreev reflection
spectra can be fitted using two fitting parameters, $Z$ and $\Delta$, the superconducting energy gap. For all non-zero values of $Z$, a $dI/dV$ spectrum shows a double-peak structure symmetric
about $V$ = 0 -- this is a hallmark of Andreev reflection. 

When the metal in the metal-superconductor point-contact
is a ferromagnet with finite spin polarization at the Fermi level, the density of states of the up-spin
electrons ($N_{\uparrow}$ ($E_F$)) is not equal to the density of states of the spin
down electrons ($N_{\downarrow}$($E_F$)). Therefore, $|N_{\uparrow}$ ($E_F$)-$N_{\downarrow}$($E_F$)$|$
electrons encountering the interface cannot undergo Andreev
reflection because they do not find accessible states in the
opposite spin band. Therefore, in a point-contact between a
ferromagnetic conductor and a conventional superconductor,
Andreev reflection is suppressed. Clearly, for a half metallic ferromagnet, the Andreev reflection is expected to be completely suppressed leading to zero conductance below $V = \pm \Delta/e$, and no features associated with Andreev reflection should appear in a $dI/dV$ vs. $V$ spectrum. For non-half metals, by measuring the degree
of the suppression of Andreev reflection, the spin-polarization of the Fermi level
can be measured. In order to extract the absolute value of
the Fermi level spin polarization, first the BTK current is
calculated for zero spin polarization ($I_{BTKu}$) and 100\% spin
polarization ($I_{BTKp}$), respectively. Then the current for an intermediate spin polarization $P_t$ is calculated by interpolation
between ($I_{BTKu}$) and ($I_{BTKp}$) following the relation $I_{total} =  I_{BTKu} (1-P_t) + P_tI_{BTKp}$. The derivative of $I_{total}$ with
respect to $V$ gives the modified Andreev reflection spectrum
with finite spin polarization in the non-superconducting electrode forming the point contact\cite{Soulen}. This model is traditionally used
to analyze the spin-polarized Andreev reflection spectra
obtained between a ferromagnetic metal and a superconductor by using four fitting parameters $Z$, $\Delta$, $\Gamma$ and $P_t$.
\begin{figure} [h!]
	\centering
		\includegraphics[width=.5\textwidth]{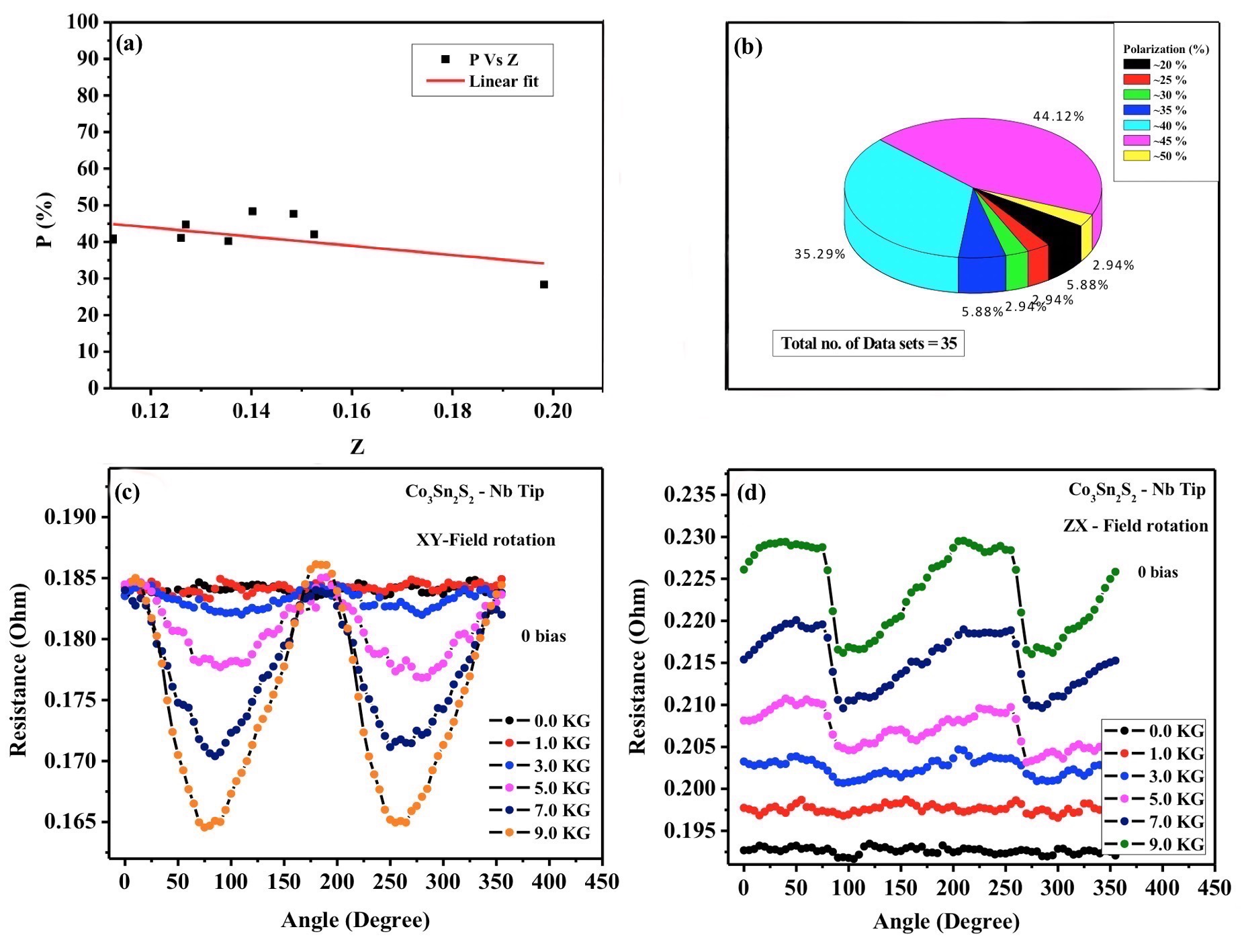}
     	\caption{ (a) Spin polarization(P) Vs barrier strength (Z) plot. Extrapolation to Z=0 is represented by solid line where spin-polarization approaches 45\% (b) statistics of spin polarization obtained for different point contacts (c),(d) Angular  magnetoresistance performed at zero voltage bias and field rotated in plane  and out of plane(perpendicular) of S/F interface respectively.  }	

\end{figure}

In Figure 1, we show a number of differential conductance ($dI/dV$) spectra obtained for a point contact between Co$_3$Sn$_2$S$_2$ and a sharp tip of the conventional superconductor Nb at varying temperatures (a) and magnetic fields (b). The two peaks in conductance around $V$=0 are the peaks due to Andreev reflection. This is surprising since Co$_3$Sn$_2$S$_2$ is known to be a half metallic ferromagnet for which Andreev reflection must be completely suppressed. To note, the conductance at zero-bias is large confirming non-vanishing Andreev reflection at Co$_3$Sn$_2$S$_2$/Nb interfaces. The spectra systematically evolve with temperature and all the features associated with Andreev reflection disappear around 9 K, the critical temperature of Nb. Similarly the features gradually disappear with increasing magnetic field, as expected.
 \begin{figure} [h!]
	\centering
		\includegraphics[width=.5\textwidth]{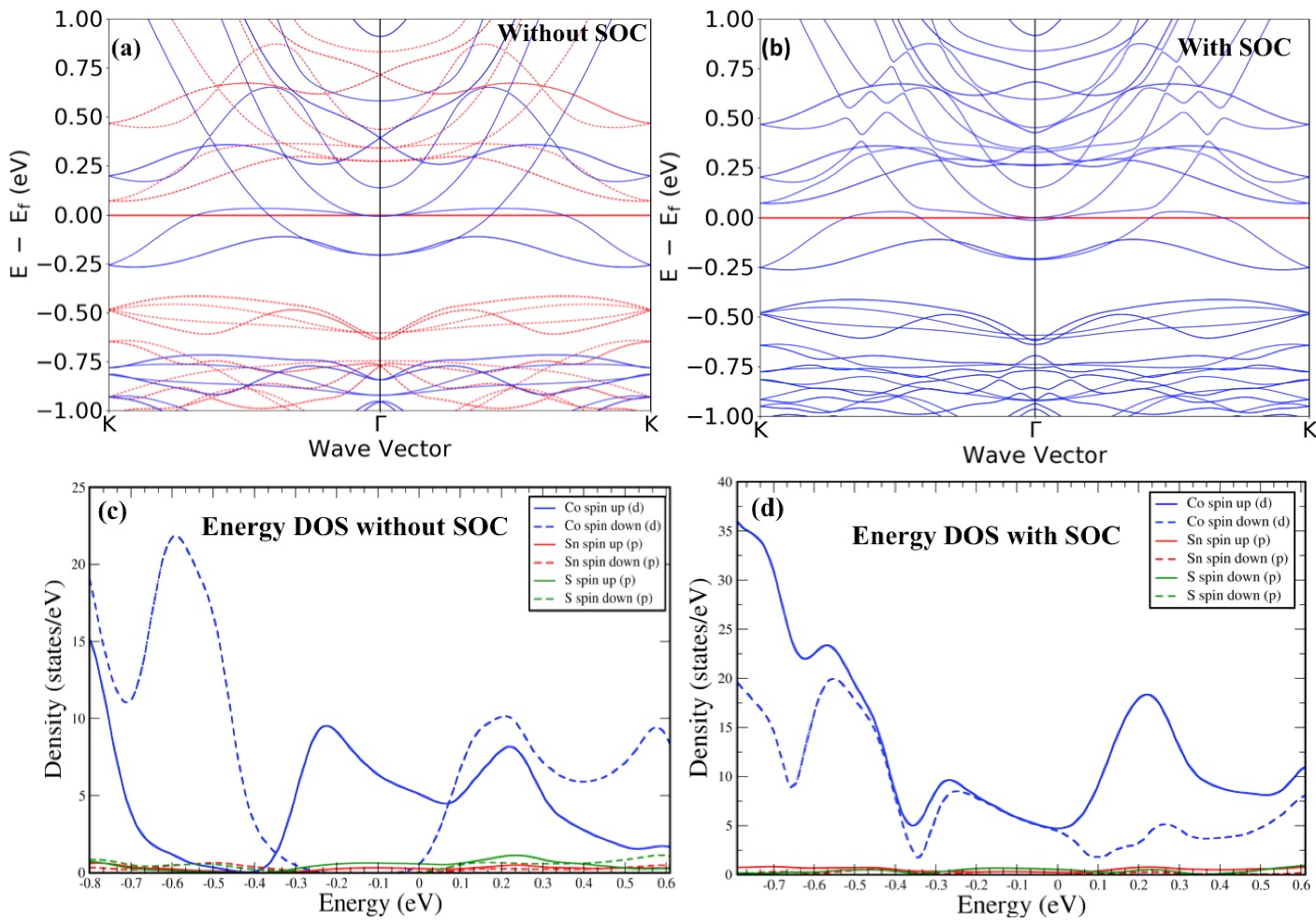}
     	\caption{ (a),(b) Theoretical bandstructure calculated without and with Spin-Orbit coupling (Red and blue color lines represent spin-up and spin-down states) (c),(d) Energy density of states calculated at Fermi level without and with Spin-Orbit coupling (In (d), we observe an overlap of spin-up and spin-down states just below the Fermi level) }	

\end{figure}

One important feature of all the spectra is an asymmetry about $V$=0. This asymmetry is always seen for this system.  Above the critical temperature and magnetic field, where the spectral features disappear, the asymmetry persists. As we will show later, this asymmetry stems from an asymmetry in the density of states when spin-orbit coupling is included in the calculation. For the analysis of spin polarization within the modified BTK model, we have extracted the symmetric component of spectra\cite{Mehta}. We have shown the temperature (a) and magnetic field (b) evolution of the point contact spectra in Figure 2. The data obtained for each temperature and magnetic field is normalized to the conductance at the highest voltage bias (the normal state). The colored points show the (symmetrized) experimental data points and the black lines show the theoretical fits within modified BTK model as discussed above. The measured superconducting energy gap ($\Delta$) is found to be 1 meV which is smaller than the gap of pristine Nb. This reduction in $\Delta$ of Nb can be attributed to the proximity of the ferromagnetic Co$_3$Sn$_2$S$_2$. However, the reduced gap evolves with temperature following the prediction within BCS theory\cite{Gasperovic,Bardeen} (Figure 3(c)). Hence, presence of the Weyl semimetal did not cause a deviation from the conventional nature of superconductivity. $\Delta$ also decreases gradually with increasing magnetic field (Figure 3(d)). The fittings presented here also provide a quantitative estimate of the  intrinsic ($Z$ = 0) transport spin polarization. In order to find this quantity, we first captured a number of spectra with varying $Z$ and analysed them. The extracted values of $P_t$ from such analysis were plotted as a function of $Z$. The extrapolation of that to $Z$ = 0 gives the intrinsic $P_t$. \cite{note} Intrinsic $P_t$ thus measured is found to be $\sim$ 50\% at lowest temperature and zero magnetic field. For the fittings up to 9 K, $P_t$ was kept constant at it's low temperature value. The distribution of $P_t$ that were used for this analysis is shown in a pie chart (Figure 3(b)). As it is seen, for none of the point contacts the spin polarization approaches 100 \%.

The presence of spin polarized current in the point contacts is further confirmed by field angle dependence of resistance of a ballistic point contact. We rotated the magnetic field using a 3-axis vector magnet with respect to the direction of applied current. The data for $V$=0 is presented in Figure 3(c: for field rotated in plane and d: for field rotated out of plane). We found that above 3 KG pronounced anisotropy arises in the field-angle dependent resistance. The anisotropy gets stronger with increasing magnetic field strength.  The anisotropy can be explained if we consider the micro-constriction to have the shape of a nano-wire and the field is rotated with respect to direction of flow of current through this wire. This kind of AMR is usually seen in materials which have surface states with complex spin structure\cite{Shang,Kandala}.      
    
Provided the fact that the measured transport spin polarization is far less than 100\%, it is imperative to understand the possible mechanism that leads to the reduction of spin polarization (spin ``unpolarization"). In order to gain understanding on that,  we carried out electronic structure calculation in the ferromagnetic phase of Co$_3$Sn$_2$S$_2$ with and without SOC. The point symmetry group of Co$_3$Sn$_2$S$_2$ unit cell is ${\bar{3}m}$. Ab-initio calculation was done using Vienna Ab-initio Simulation Package(VASP)\cite{vasp4}. To describe the core electrons we have used projector augmented wave(PAW)\cite{PAW} pseudo-potentials and for exchange-correlation functional we have used Perdew-Burke-Ernzerhof (PBE)\cite{PBE} form. The cut-off energy for plane wave basis was set to  $500$ $\mathrm{eV}$. The Monkhorst-Pack k-mesh was set to $15\times15\times6$ in the Brillouin zone for the self-consistent calculation. The unit cell for ${R\bar{3}m(166)}$ space group can be represented as an hexagonal or a rhombohedral representation. We have used hexagonal representation. We have used the experimentally obtained lattice parameters.\cite{Exp_latt_Cons}

In Figure 4, we present our band structure and DOS calculations without (left panel) and with (right panel) SOC. Without SOC, we find large spin splitting in the band structure as well as in the DOS with a full spin-polarization at the states near the Fermi level. Expectedly, the low-energy states are dominated by the Co-$d$ orbitals. Although the SOC strength of the Co-$d$ is often expected to be less, here due to the presence of Sn and S cations and owing to the loss of time-reversal symmetry in the ferromagnetic state, the spin-orbit splitting is enhanced. Although the band structure shows similar dispersion features as in the case without the SOC, the DOS shows a dramatic and interesting spin {\it depolarization} with SOC. In light of the experimental data, the key results to highlight here are as follows. (i) We find that the spin-polarization is reversed {\it above} the Fermi level with SOC. While spin-down channels had more states than the spin-up channels without SOC, the spin-up states become largely populated after the SOC is introduced. (ii) Just {\it below} the Fermi level, both spin up and spin down states become {\it equally} populated upto the binding energy of 250~meV. This result is fully consistent with the experimental observations of  50\% spin-polarization in terms of Andreev reflection. (iii) While the DOS itself is slightly particle-hole asymmetric, the spin-polarization has considerably lost this symmetry. This delineates the source of particle-hole asymmetry as observed in the experimentally obtained Andreev reflection spectra .     
     
In conclusion, from spin-polarized Andreev reflection spectroscopy on the ferromagnetic Weyl semimetal Co$_3$Sn$_2$S$_2$ we have shown that the transport spin polarization in the system is far less than what is expected for a half metal. In addition, the Andreev  reflection spectra are reproducibly asymmetric regardless of the barrier strength. Our calculations reveal that both the key observations namely the spin depolarization and the anisotropic Andreev reflection spectra are consequence of spin-orbit coupling which leads to equal population of the spin up and the spin down bands near the Fermi level.

RK acknowledges CSIR, India for senior research fellowship. GS acknowledges financial support from an SERB extramural grant (grant number: \textbf{EMR/2016/003998}).


\begin{thebibliography}{100}
\bibitem{note} To note, there is no physical reason why the extrapolation has to be linear, but that is what is typically done in modified BTK analysis. Nevertheless, the main result namely the observation of non-half metallicity is not altered by this.
\end{thebibliography}
\end{document}